\documentclass{ptapap}

\author{G.A. Wade}[RMC]
\author{D.H. Cohen}[Swarthmore]
\author{C. Fletcher}[FIT] 
\author{G. Handler}[NCAC] 
\author{L. Huang}[RMC] 
\author{J. Krticka}[Brno] 
\author{C. Neiner}[Obspm] 
\author{E. Niemczura}[Wroclaw] 
\author{H. Pablo}[UdeM] 
\author{E. Paunzen}[Brno] 
\author{V. Petit}[FIT] 
\author{A. Pigulski}[Wroclaw] 
\author{Th. Rivinius}[ESO] 
\author{J. Rowe}[UdeM] 
\author{M. Rybicka}[NCAC] 
\author{R. Townsend}[Wisc] 
\author{M. Shultz}[Queens] 
\author{J. Silvester}[Uppsala] 
\author{J. Sikora}[Queens]
\author{the BRITE-Constellation Executive Science Team (BEST)}
\affil[RMC]{Dept. of Physics, Royal Military College of Canada, Canada}
\affil[Swarthmore]{Dept. of Physics and Astronomy, Swarthmore College, USA}
\affil[FIT]{Dept. of Physics and Space Sciences, Florida Institute of Technology, USA}
\affil[Wisc]{Dept. of Astronomy, University of Wisconsin-Madison, USA}
\affil[NCAC]{Copernicus Astronomical Center, Poland}
\affil[Brno]{Dept. of Theoretical Physics and Astrophysics, Masaryk University, Czech Republic}
\affil[Obspm]{LESIA, Observatoire de Paris, PSL Research University, CNRS,
Sorbonne Universit\'es, UPMC Univ. Paris 6, Univ. Paris  Diderot, Sorbonne
Paris Cit\'e, 5 place Jules Janssen, 92195 Meudon, France}
\affil[UdeM]{D\'ept. de Physique, Universit\'e de Montr\'eal, Canada}
\affil[Queens]{Dept. of Physics, Engineering Physics and Astronomy, Queen's University, Canada}
\affil[Wroclaw]{Astronomical Institute, University of Wroc\l{}aw, Kopernika 11, 51-622 Wroc\l{}aw, Poland}
\affil[ESO]{European Southern Observatory, Chile}
\affil[Uppsala]{Uppsala University, Sweden}

\title{Magnetic B stars observed with BRITE: Spots, magnetospheres, binarity, and pulsations}
\headtitle{Magnetic B stars observed with BRITE}

\begin{document}

\maketitle

\begin{abstract}

Magnetic B-type stars exhibit photometric variability due to diverse causes,
and consequently on a variety of timescales. In this paper we describe interpretation of 
BRITE photometry and related ground-based observations of 4 magnetic
B-type systems: $\epsilon$~Lupi, $\tau$~Sco, a~Cen and $\epsilon$~CMa.

\end{abstract}

\section{Introduction}

Approximately 10\% of mid- to early-B stars located on the main sequence show direct evidence of strong surface magnetism. Studying the photometric variability of such systems provides insight into their multiplicity and physical characteristics, rotation, surface and wind structures, and pulsation properties. For example, in late- and mid- B-type stars (below about spectral type B2) magnetic fields stabilise atmospheric motions and allow the accumulation of peculiar abundances (and abundance distributions) of various chemical elements. At earlier spectral types, magnetic fields channel radiatively-driven stellar winds, confining wind plasma to produce complex co-rotating magnetospheres. Some magnetic stars are located in close binary systems, where photometric variability may reveal eclipses, tidal interaction and (potentially) mass and energy transfer effects. Finally, some magnetic B stars are located in an instability strip, and exhibit $\beta$~Cep and SPB-type pulsations. 

The bright magnetic B stars observed by the BRITE-Constellation have been preferential targets of spectropolarimetric monitoring within the context of the BRITEpol survey (see the paper by Neiner et al. in these proceedings). In this article we provide brief reports on analysis of BRITE photometry and complementary data for 4 magnetic B-type stars for which the BRITE observations detect or constrain variability due to these mechanisms.

\section{$\epsilon$ Lupi}

\begin{figure}
\includegraphics[width=\textwidth]{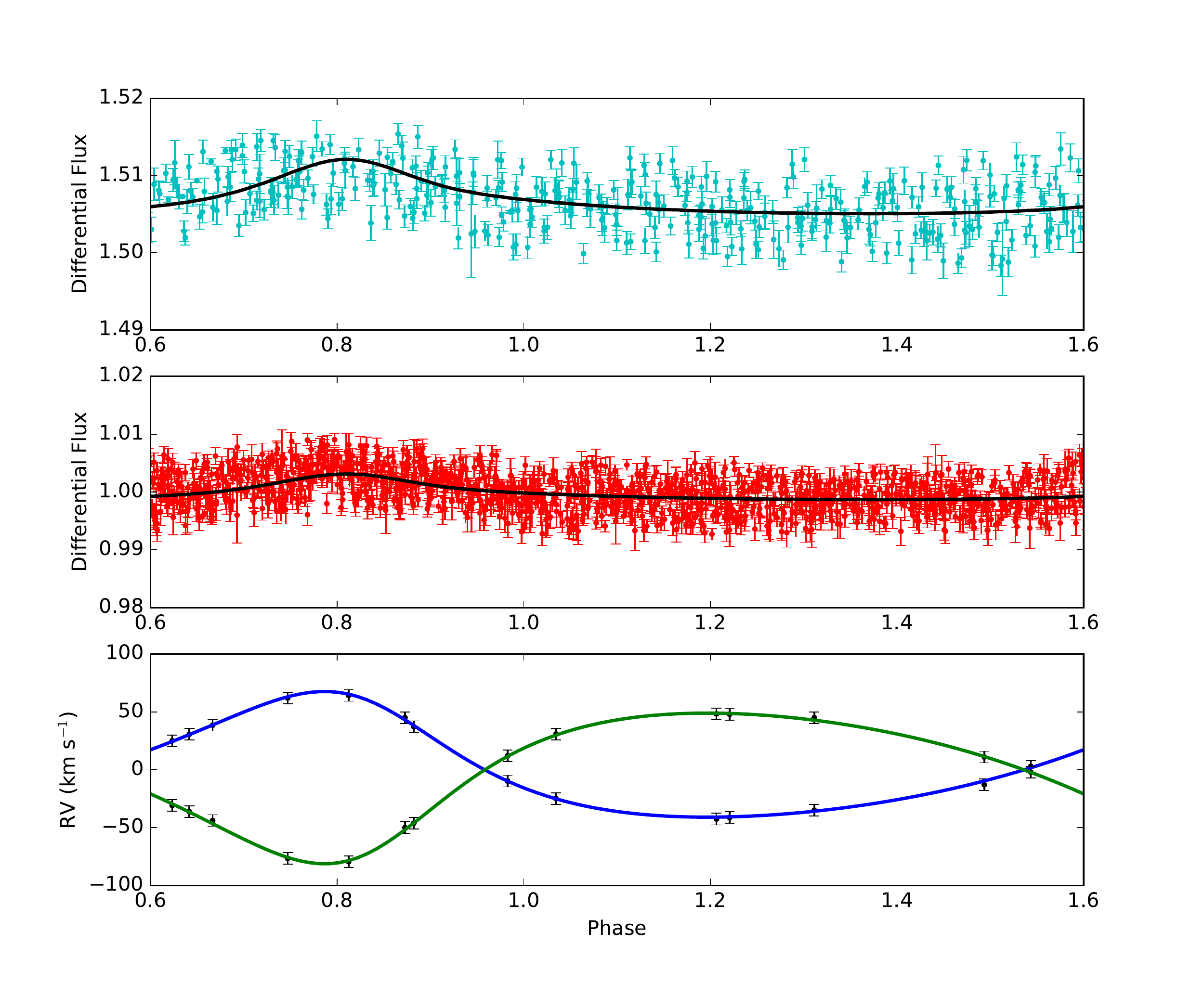}
\caption{Phased photometry (Upper panel - BRITE blue, middle panel - BRITE red) and radial velocities (lower panel) of $\epsilon$~Lupi, compared with the predictions of the heartbeat model (curves).}
\label{fig:epslupi}
\end{figure}

$\epsilon$~Lupi is a short-period ($\sim 4.6$~d) eccentric binary system
containing two mid/early-B stars \citep[B3V/B2V, ][]{2005A&A...440..249U}.
It was observed by the BRITE UBr, BAb, BTr, BLb nano-satellites \citep[see e.g.][]{2016PASP..128l5001P} during the Centaurus campaign from
March to August 2014, and again by BLb during the Scorpius campaign from February to August 2015.

{Magnetic fields associated with both the primary and secondary components
were reported by \citet{2015MNRAS.454L...1S}, making $\epsilon$~Lupi the
first known doubly-magnetic massive star binary}. The (variable) proximity
of the two components led  \citet{2015MNRAS.454L...1S} to speculate that
their magnetospheres may undergo reconnection events during their orbit.
Such events, as well as rotational modulation by surface structures and
the suspected $\beta$~Cep pulsations of one or both components, could
introduce brightness fluctuations potentially observable by BRITE.

The periodogram of the BRITE photometry shows power at the known orbital
period. When the data are phased accordingly, both the red (BTr+UBr) and
blue (BAb) lightcurves exhibit a subtle, non-sinusoidal modulation with
peak flux occurring at the same phase as the orbital RV extremum (i.e.
periastron). We interpret this modulation as a ``heartbeat" effect \citep[e.g.][]{2012ApJ...753...86T},
resulting from tidally-induced deformation of the stars during their close passage at periastron.
Assuming this phenomenon, we have successfully modeled the lightcuves and RV variations using the PHOEBE code \citep[version 1,][see
Fig.~\ref{fig:epslupi}]{2005ApJ...628..426P}.

\begin{figure}
\includegraphics[width=\textwidth]{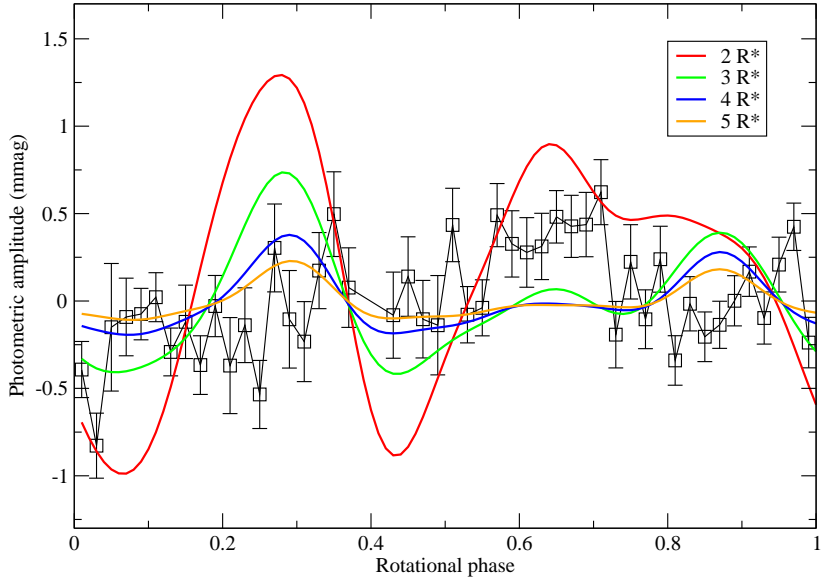}
\caption{BRITE red filter photometry of $\tau$~Sco, compared with the predictions of ADM models computed assuming pure scattering. The different colours correspond to source surface radii ranging from 2-5~$R_*$.}
\label{fig:tausco}
\end{figure}

\section{$\tau$~Sco}

$\tau$~Sco is a hot main sequence B0.5V star that was observed by BAb, UBr, BLb, BHr during the Scorpius campaign from February August 2015. \citet{2006MNRAS.370..629D} detected a magnetic field in the photosphere of this X-ray bright star, varying according to a rotational period of 41~d. They modeled the magnetic field topology, finding it to be remarkably complex. \citet{2010ApJ...721.1412I} acquired Suzaku X-ray measurements of $\tau$~Sco. They found that the very modest phase variation of the X-ray flux was at odds with the predicted variability according to the 3D force-free extrapolation of the magnetosphere reported by \citet{2006MNRAS.370..629D}.

Petit et al. (in prep) have sought to explain this discrepancy by reconsidering the physical scale of the closed magnetospheric loops of $\tau$~Sco. New modeling of system using the Analytic Dynamical Magnetosphere (ADM) formalism \citep{2016MNRAS.462.3830O} yields predictions of the X-ray variability as a function of the adopted mass-loss rate (as quantified by the ``source surface" of the extrapolation).

These same ADM models have been used in conjunction with BRITE photometry to constrain the distribution of cooler plasma surrounding the star. Adopting a pure electron scattering approximation, we have computed the expected brightness modulation as a function of source surface distance (Fig.~\ref{fig:tausco}). The very high quality of the BRITE red photometry allows us to rule out models with source surface radii smaller than 3~$R_*$.

\section{a~Cen}

a~Cen is a Bp star of intermediate spectral type ($T_{\rm eff}\sim 19$~kK) that exhibits extreme variations of its helium lines during its 8.82~d rotational cycle. It was observed during the Centaurus campaign from March to August 2014 by UBr, BAb, BTr, BLb. 

\citet{2010A&A...520A..44B} used high resolution spectra to compute Doppler Imaging maps of the distributions of He, Fe, N and O of a Cen, revealing in particular a more than two-order-of-magnitude contrast in the abundance of He in opposite stellar hemispheres. They also discovered that the He-poor hemisphere shows a high relative concentration of $^3$He. 

\begin{figure}
\includegraphics[width=6.5cm]{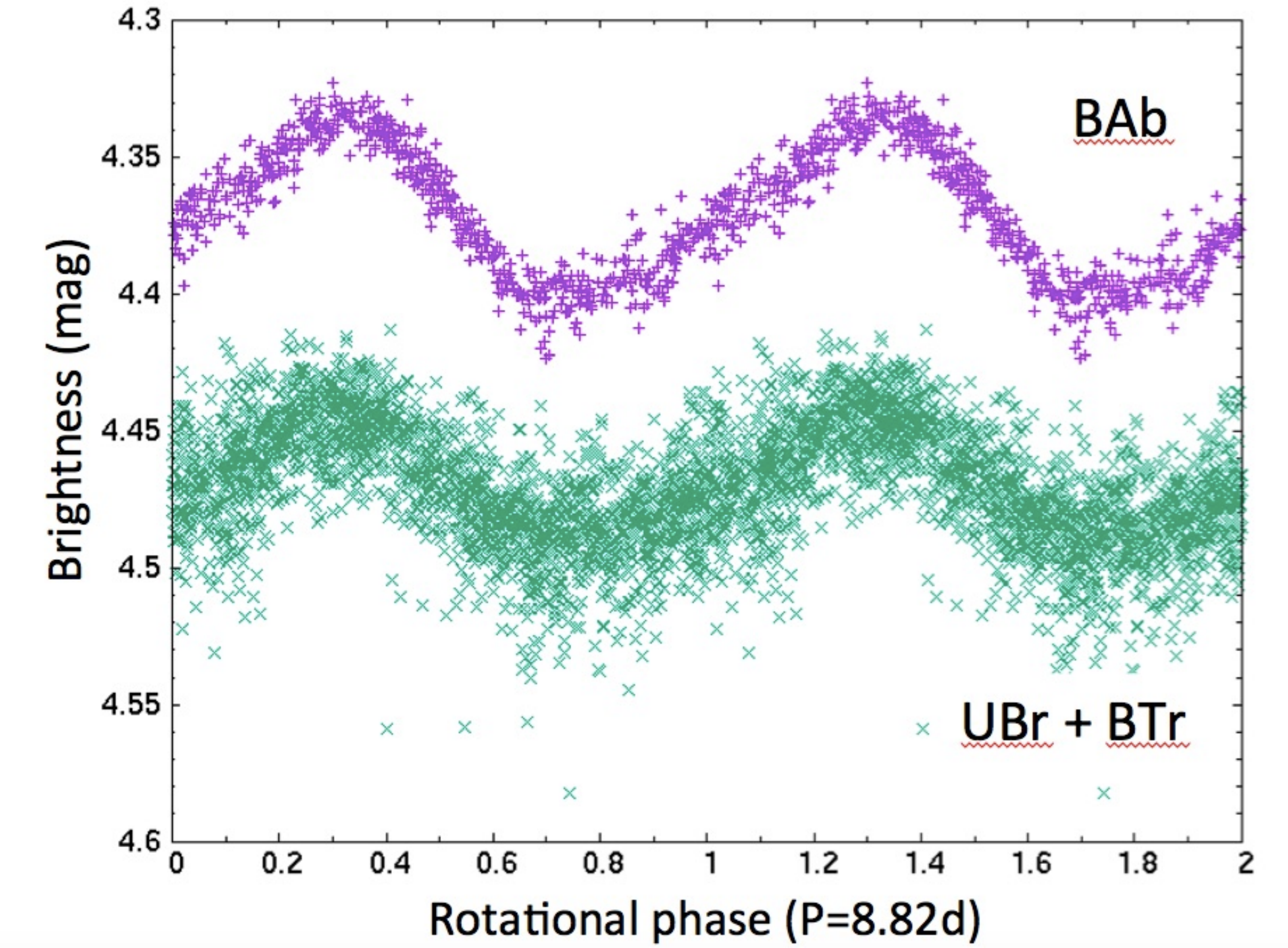}\includegraphics[width=6.5cm]{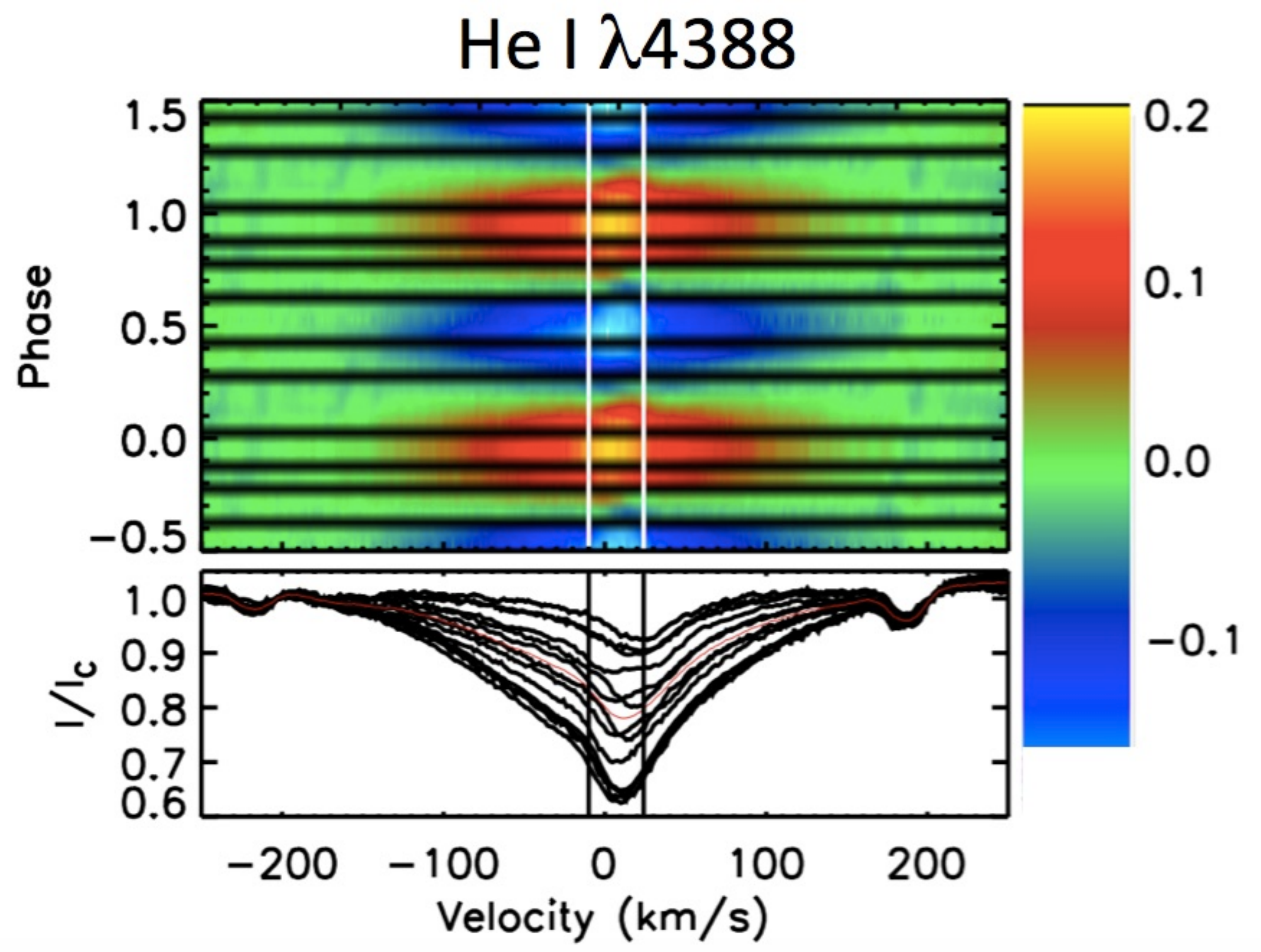}
\caption{{\em Left panel -}\ BRITE photometry of a Cen (upper curve - blue filter, lower curve - red filter) phased according to the stellar rotation period. {\em Right panel -}\ Dynamic spectrum of the rotational variability of the He~{\sc i}~$\lambda 4388$ line, showing the extreme changes in line strength.}
\label{fig:acen1}
\end{figure}

\begin{figure}
\includegraphics[width=\textwidth]{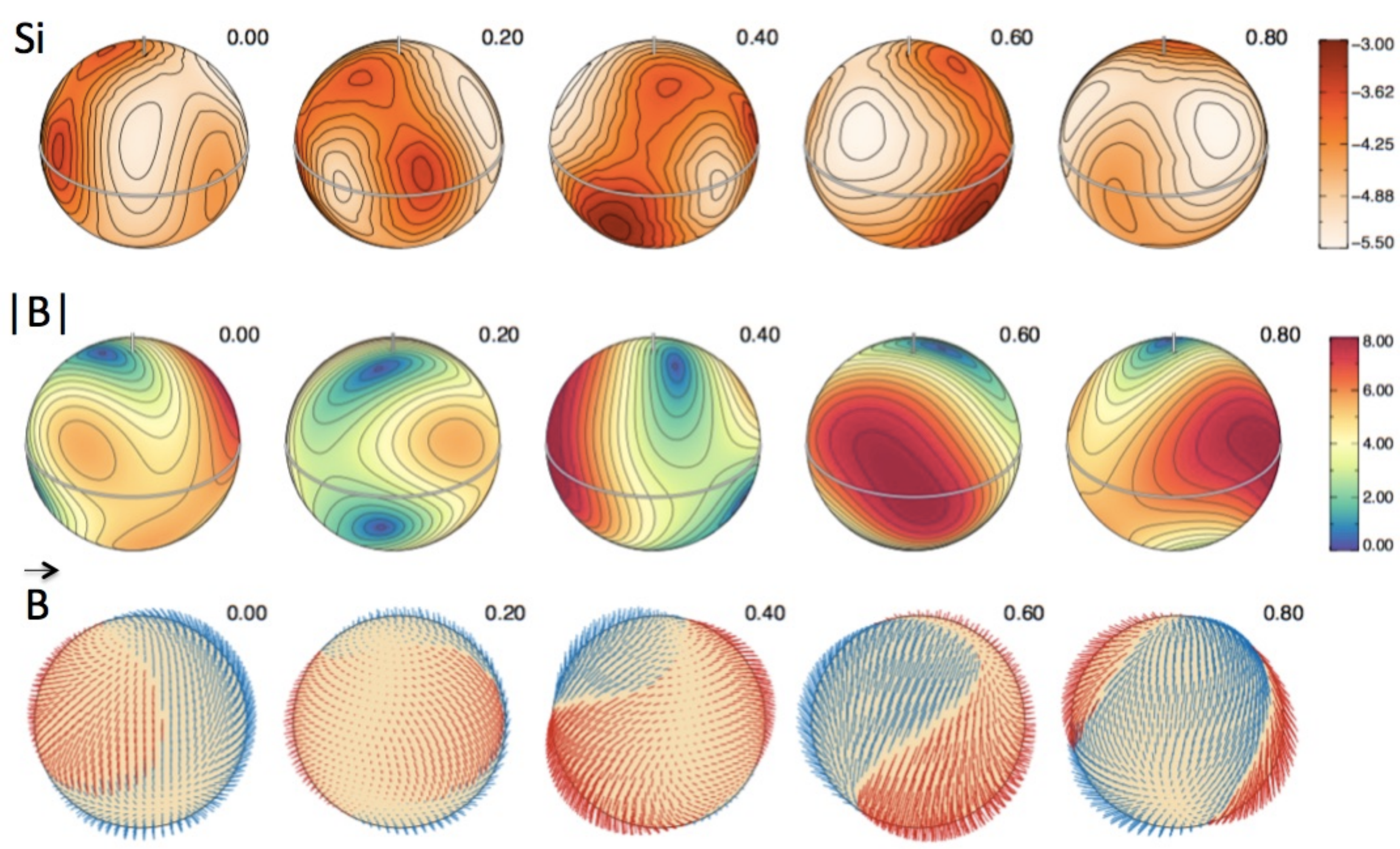}
\caption{Magnetic Doppler Imaging maps of a Cen, showing the surface Si distribution (upper row), and the magnetic field modulus and orientation (middle and bottom rows, respectively). }
\label{fig:acen2}
\end{figure}

The BRITE photometry of a Cen exhibits clear variability according to the previously-known rotational period (Fig.~\ref{fig:acen1}, left panel). It also reveals marginal variability at frequencies that may correspond to pulsations in the SPB range. Using a collection of 19 new ESPaDOnS and HARPSpol Stokes $V$ spectra, in addition to archival UVES spectra (e.g. Fig.~\ref{fig:acen1}, right panel), new self-consistent Magnetic Doppler Imaging maps have been derived of the stellar magnetic field and the abundance distributions of various elements, including Si (Fig.~\ref{fig:acen2}). These maps will be used as basic input for modeling the two-colour BRITE lightcurves \citep[e.g.][]{2009A&A...499..567K}.

\section{$\epsilon$~CMa}

$\epsilon$~CMa is an evolved B1.5II star. It was observed during the Canis Majoris-Puppis campaign from October 2015 to April 2016 by UBr, BLb, BTr. A weak magnetic field was detected in photospheric lines of this star by \citet{2015A&A...574A..20F}. 

\begin{figure}
\includegraphics[width=\textwidth]{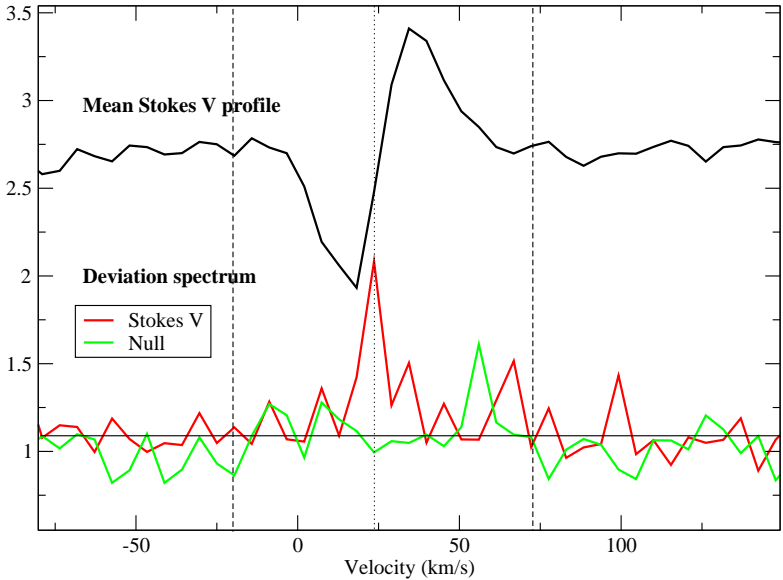}
\caption{Results of the variability analysis of over 120 LSD Stokes $V$ profiles of $\epsilon$~CMa, showing the mean profile (shifted vertically for display purposes) and per-pixel deviation of both Stokes $V$ and the diagnostic null ($N$) normalised to their respective uncertainties. The dashed lines show the velocity span of the $V$ profile and its centre-of-gravity. The largest deviation of $V$ occurs near the centre-of-gravity of the profile, and corresponds to only 2$\sigma$.}
\label{fig:epscma}
\end{figure}

A preliminary analysis of the BRITE photometry reveals no significant variability.  However, we have continued to monitor the magnetic field of $\epsilon$~CMa, analysing over 120 Stokes $V$ exposures obtained over a span of 125~d, with the aim of (i) detecting rotational modulation and determining the stellar rotational period, and (ii) modeling the surface magnetic field strength and geometry. Variability of the Stokes $V$ profiles - as quantified by a deviation analysis (Fig.~\ref{fig:epscma}) - is very weak. At the level of precision of the magnetic data (best error bars of 2~G, median error bar of 4~G), no periodic variability can be inferred. Considering the reported projected rotational velocity and measured angular diameter of the star, the rotational period should be no longer than $\sim 25$~d. This could imply that the star is viewed close to the rotational pole, that the magnetic axis is aligned with the rotation axis, or that the global field contrast is significantly weaker than expected from a dipole.

\acknowledgements{CN thanks PNPS for its support to the BRITE spectropolarimetric program. EN is supported by NCN through research grant No. 2014/13/B/ST9/00902. RHDT acknowledges support from NSF SI2 grant ACI-1339600 and NASA TCAN grant NNX14AB55G. DHC acknowledges Chandra awards TM4-15001B and AR2-13001A. CF and VP acknowledge NASA grant NNX15AG33G. The Polish contribution to the BRITE mission is funded by the BRITE PMN grant 2011/01/M/ST9/05914. GH and MR acknowledge support by NCN grant 2015/18/A/ST9/00578. GAW acknowledges Discovery Grant support from the Natural Sciences and Engineering Research Council (NSERC) of Canada. GAW and HP acknowledge Andrew Tkachenko for a clever observation made by him during his presentation at the second BRITE Science Meeting.}

\bibliographystyle{ptapap}
\bibliography{wade}

\end{document}